\documentclass{article}
\usepackage{anysize,url,cite}
\marginsize{2.3cm}{2.3cm}{2.3cm}{2.3cm}
\title{A review of ``Memcomputing NP-complete problems
in polynomial time using polynomial resources'' (\url{arXiv:1411.4798})
}
\author{Igor L. Markov}
\date{}

\begin{document}
\maketitle
\abstract{
The reviewed paper describes an analog device that empirically solves small instances of the NP-complete Subset Sum Problem (SSP). The authors claim that this device can solve the SSP in polynomial time using polynomial space, in principle, and observe no exponential scaling in resource requirements. We point out that (a) the properties ascribed by the authors to their device are insufficient to solve NP-complete problems in poly-time, (b) runtime analysis offered does not cover the spectral measurement step, (c) the overall technique requires exponentially increasing resources when scaled up because of the spectral measurement step.
}

\ \\
{\bf 1.} The paper reports a physical device and claims that this device solves a particular NP-complete problem using only polynomial time and space. The idea is elegant - $n$ input parameters are entered as frequencies of periodic signals, and combined in such a way that the spectrum of the system represents partial and full sums of those parameters. Rather than iterate through this exponentially-sized spectrum, the authors develop a direct check if a particular value appears there. This solves the Subset-Sum Problem - checking a particular value can be presented by adding up some of the provided values. The operation of the device is illustrated on small examples $n=4,5,6$, and the results are credible. The principle of operation is simple enough not to require mathematical proofs. The authors are careful enough to point out that their work does not imply that P=NP because the proposed machine is more appropriately modeled by a non-deterministic Turing machine, so it is hardly surprising that it solves NP-complete problems. The surprising claim here is the possibility of a physical realization of a non-deterministic Turing machine (or an equivalent), although this seems to have already been discussed by the authors in Ref. [2] of the reviewed paper. Perhaps, the novelty is the specific empirical demonstration.

\ \\
\noindent
{\bf 2.} The most significant consequence of this work would be the disproof of the Physical Church-Turing thesis \cite{Ben-Amram}, which states that {\em a deterministic Turing machine can efficiently simulate any physically-realistic model of computation} (except possibly for some models relying on quantum physics). Surprisingly, the paper never mentions this. The thesis was important enough to have been heavily studied since the 1940s by the best minds. In the 1940s, Ulam and von Neumann studied a (non-von-Neumann, in modern terms) model of computation called cellular automata \cite{CA}. Among its advantages, we find ``{\bf intrinsic parallelism - interacting memory cells simultaneously and collectively change their states when performing computation}” (here I am quoting from the reviewed paper the first property of their new device). Cellular automata gave rise to Field- Programmable Gate Arrays (FPGAs) that one can find in some modern consumer electronics. FPGAs enjoy ``{\bf functional polymorphism – depending on the applied signals, the same interacting memory cells can calculate different functions}'' (here I am quoting from the reviewed paper the second property of their new device). While faster than single CPUs in some cases, FPGAs are polynomial-time equivalent to Turing machines.

\ \\
\noindent
{\bf 3.} In the 1970s, when NP-completeness was developed by Cook, it was noted that unbounded-precision arithmetic is sufficient to solve some difficult problems quickly \cite{Schonhage,Shamir}. By the same token, analog computing with unlimited precision gives power to solve some hard problems. Digital arithmetic with very large precision is not prohibited by fundamental physical limits (digital circuits scale out well), whereas ultra-precise analog measurements often conflict with Planck's constant, the Heisenberg uncertainty principle, the Pauli exclusion principle, and various derived results. In engineering terms, analog devices are limited because they cannot effectively counter noise (another reason is discussed below). But theoretical models of computation where analog quantities enjoy unbounded precision are questionable as well.  In this context, the analog arithmetic used in the proposed new device, and its very high sensitivity to noise should be seen as early warning signs.

\ \\
\noindent
{\bf 4.} In the 1980s, Richard Feynman studied the Physical Church-Turing thesis in its extended form, with quantum physics included. The main advantage here is that ``{\bf a group of interacting memory cells can store a quantity of information which is not simply proportional to the number of memory cells itself}'' (here I am quoting from the reviewed paper the third and last property of their new device, despite the unfortunate wording) - in fact, $n$ quantum bits can take on a quantum state with $2^n$ different amplitudes \cite{QCQI}. However, in the 1990s E. Bernstein and U. Vazirani proved \cite{QT} that quantum Turing machines do not offer super-polynomial advantage on NP-complete problems compared to conventional deterministic Turing machines.

\ \\
\noindent
{\bf 5.} The first property pitched in the reviewed paper --- {\em parallelism} --- is, of course, well studied. A 1988 result by David Fisher \cite{Fisher} analyzes physically-realistic parallelism where the size of computational units is lower- and upper-bounded by constants, while the number of dimensions and the communication speed are upper-bounded. Fisher's result points out that (without quantum physics) exponential speed-up from parallelism is impossible with even an unbounded supply of units, as long as the outputs actually depend on the inputs (obviously, it is possible to invert $N$ bits in parallel with exponential speed-up, but that would be $N$ separate computations).

The second property --- {\em reconfigurable units} --- gives no advantage in the context of poly-time equivalence of Turing machines, since each Turing machine can perform any function.

The third property - {\em being able to store in $N$ memory cells more than the sum of information stored in individual cells} - is indeed attractive and reminds of quantum entanglement \cite{QCQI}. Three issues arise with this claim.
The authors configure those memory cells individually by setting $N$ numbers, so claims of storing a greater amount of information seem suspect - regardless of how information is processed, no new information is created after input.
As in the case of quantum entanglement, the readout creates many difficulties (the Holevo bound gives a hard limit on how much information can be extracted from a quantum superposition \cite{QCQI}).
Quantum circuits enjoy the three properties listed as the advantages of the proposed device, and in fact rely on spectral techniques, but provably do not help with NP-complete problems.
Intuitively, the proposed device is somewhere between conventional and quantum computing, but since the two are equally powerful on NP-complete problems (up to poly factors), the proposed device should not offer anything new. So, what did the authors slide under the rug, aside from unbounded-precision analog computation?

\ \\
\noindent
{\bf 6.} The material on both sides of the section title ``SOLVING THE SSP USING POLYNOMIAL RESOURCES'' is worth particular attention. Before the title, we are given energy analysis of the proposed device. However, after the title, we are reminded that one additional step is needed - read-out. The full spectrum is too big to be read value-by-value, but ``{\bf a solution to this problem can be found by just using standard electronics to implement a read-out unit capable of extracting the desired frequency amplitude without adding any computational burden}''. For some reason, this read-out unit seems excluded from energy and runtime analysis in the draft. However, just a single spectral line out of an exponential number of them will usually carry a very small fraction of total energy. Distinguishing the measured value from zero may require exponentially large energy-time resources, while  thermodynamic noise puts up even more tangible obstacles to measurement. Simplified models of computation that dismiss these concerns had been explored 30-40 years ago and, as is known today, misrepresent scalable computing in the physical world.

\ \\
\noindent
{\bf 7.} An exponentially wide spectrum poses another challenge. If input frequencies are spaced far enough apart to disallow exponentially-close spectral lines (to avoid measurement difficulties), the spectrum will necessarily imply exponentially long wavelengths. This may require measurement of exponentially long duration using exponentially long hardware. Note how positionless analog computation cannot capture the wide ranges of values that positional digital systems represent effortlessly.

\ \\
\noindent
{\bf 8.}
There is a long history of unsuccessful attempts to solve NP-complete problems using non-standard physical models of computation. Analysis of such attempts can be found in Scott Aaronson's extensive survey \cite{Aaronson}.


\begin{thebibliography}{99}
\bibitem{Aaronson}
S. Aaronson, ``NP-complete Problems and Physical Reality'', \url{arXiv:quant-ph/0502072}.
\bibitem{Ben-Amram}
A. M. Ben-Amram, ``The Church-Turing Thesis and its Look-Alikes,'' SIGACT News 36 (3): 113–116, 2005.
\bibitem{QT}
E. Bernstein, U. Vazirani, ``Quantum complexity theory,'' {\em SIAM Journal on Computing} 26 (5): 1411–1473, 1997.
\bibitem{Fisher}
D. C. Fisher, ``Your favorite parallel algorithms might not be as fast as you think,''
{\em IEEE Trans. Computers}, 1988.
\bibitem{QCQI}
M.~A.~Nielsen and I.~L.~Chuang, ``Quantum computation and quantum information,'' 2nd ed, 2010.
\bibitem{Shamir} A. Shamir, ``Factoring Numbers in $O(\log n)$ Arithmetic Steps,''
{\em Inf. Proc. Letters} 8(1), 1979, pp. 28–31.
\bibitem{Schonhage} A. Sch\"onhage, ``On the power of random access machines,''
{\em Int. Conf. Automata, Languages and Programming} (ICALP) 1979, pp. 520-526.
\bibitem{CA}
J. von Neumann, ``The general and logical theory of automata,'' in L.A. Jeffress, ed., {\em Cerebral Mechanisms in Behavior – The Hixon Symposium}, John Wiley \& Sons, New York, 1951, pp. 1–31.
\end{thebibliography}
\end{document}